\documentclass{PoS}

\title{SuperNEMO -- the next generation double beta decay experiment}

\ShortTitle{SuperNEMO -- the next generation double beta decay experiment}

\author{\speaker{Irina Nasteva} on behalf of the SuperNEMO collaboration\\
        University of Manchester\\
        E-mail: \email{Irina.Nasteva@manchester.ac.uk}}

\abstract{The SuperNEMO experiment is being designed to search for neutrinoless double beta decay to test if neutrinos are Majorana particles. 
The experimental technique follows that of the currently running NEMO-3 experiment,
which successfully combines tracking and calorimetry to measure the topology and energy of the final state electrons. 
Unique particle identification capabilities of SuperNEMO will be employed with about 100 kg of $^{82}$Se and will reach sensitivity to a half-life of about $2\cdot10^{26}$ years, which  corresponds to Majorana neutrino masses of about 50 meV, depending on the calculated value of the nuclear matrix element. 
In this poster, the current status of the SuperNEMO project is presented.}

\FullConference{European Physical Society Europhysics Conference on High Energy Physics,
EPS-HEP 2009,\\
		 July 16 - 22 2009\\
		 Krakow, Poland}

\begin{document}

\section{Neutrinoless double beta decay}
Neutrinoless double beta decay ($0\nu\beta\beta$) is a process beyond the Standard Model, which violates lepton number conservation. 
It is only possible if neutrinos are massive Majorana particles, i.e.~the neutrino and antineutrino are identical. 
In the Standard Model, two-neutrino double beta decay ($2\nu\beta\beta$) can occur in nuclei for which beta decay is energetically suppressed. 
The signature of neutrinoless double beta decay is a peak at the end point of the energy sum spectrum of electrons from the $2\nu\beta\beta$ process. 
SuperNEMO will detect double beta decay events by measuring the energies, trajectories and vertices of electrons. It will use particle identification of photons and alpha particles to reject backgrounds. 

\section{SuperNEMO detector design}
SuperNEMO is a next-generation $0\nu\beta\beta$ experiment based on the NEMO-3 \cite{nemo} technique of tracking and calorimetry. It will search for neutrinoless double beta decay in $\sim$100~kg of enriched isotopes, reaching a sensitivity for the half-life of $T_{1/2}= 2\cdot 10^{26}$ years, which corresponds to neutrino mass sensitivity of $\sim$50~meV~\cite{nme}. 

\begin{figure} 
\centering
\includegraphics[height=.33\textwidth]{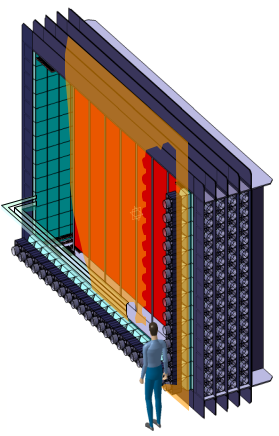}\hspace{1.5pc}
\includegraphics[height=.28\textwidth]{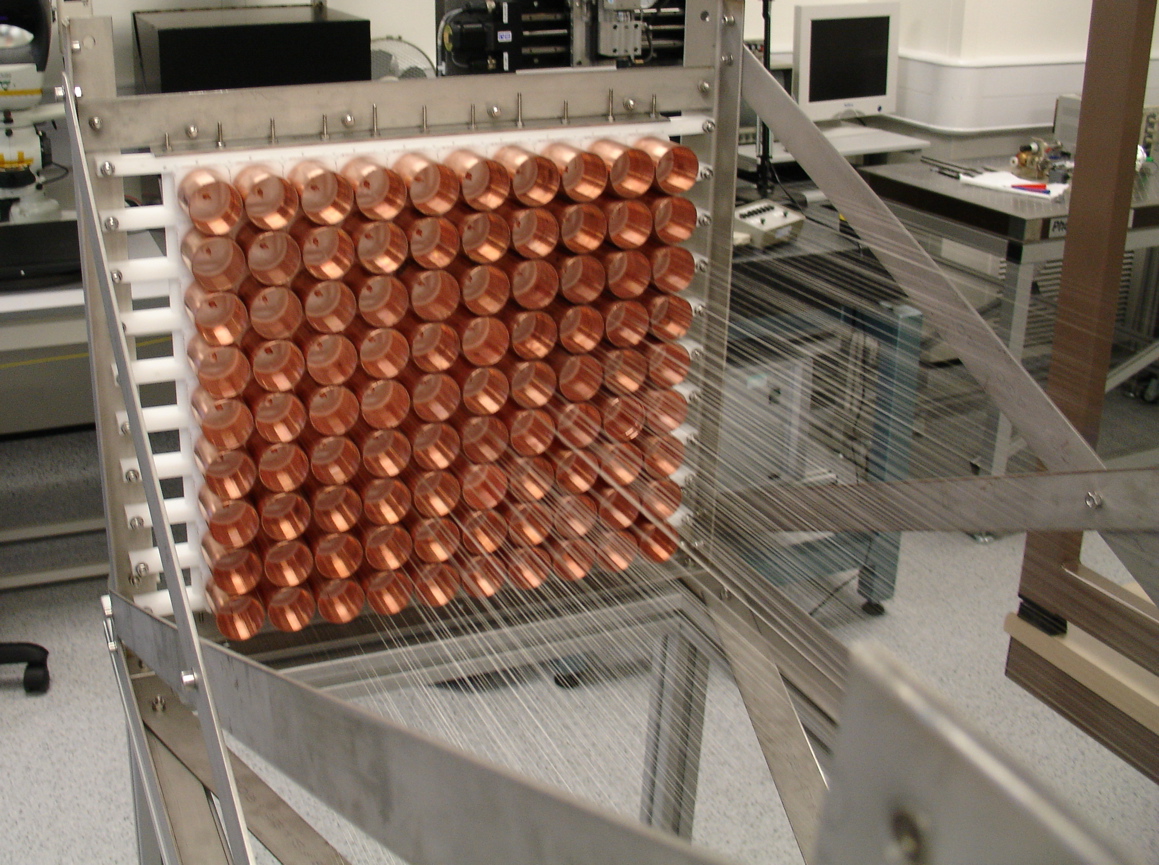}\hspace{1.5pc}
\includegraphics[height=.28\textwidth]{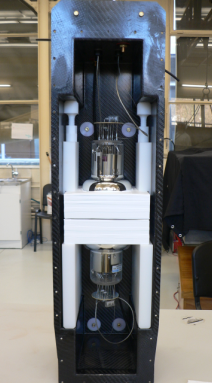}\\
(a)\hspace{13.pc}(b)\hspace{11.pc}(c)
\caption{(a) Design of a SuperNEMO module; (b) the 90-cell tracker prototype; and (c) the BiPo-1 detector.} \label{fignemo}
\label{fig1} 
\end{figure}

The SuperNEMO detector will consist of 20 identical modules, each housing 5--7 kg of isotope. 
The source is a thin foil of double beta emitter ($^{82}$Se or $^{150}$Nd). 
It is surrounded by a tracking chamber, containing drift cells operating in Geiger mode. 
A small magnetic field of $\sim$25 G is applied for electron--positron discrimination. 
The outer walls of the module make up the calorimeter, consisting of scintillators and low-radioactivity PMTs.
A three-dimensional cut-away view of a SuperNEMO module is shown in Fig.~\ref{fignemo}(a).

The SuperNEMO experiment entered a three-year R\&D phase in 2006, the outcome of which is a technical design report due in 2009. 
The construction of the first "demonstrator" module with about 5~kg of $^{82}$Se is expected to commence in 2010.
If successful, it will be followed by 19 more similar modules with anticipated completion in 2012.

\section{Calorimeter R\&D}
The energy resolution of SuperNEMO is a combination of energy losses in the foil and light collection in the calorimeter. 
The goal of the calorimeter R\&D is to achieve resolution of 7\% FWHM at 1 MeV\@. 
Two possible designs are investigated for the choice of calorimeter technology: hexagonal blocks coupled to 8'' PMTs, and 2-metre long bars coupled to 3'' PMTs at both ends.
Both use plastic PVT scintillators but differ in their operating characteristics.
The block design uses the familiar technology of NEMO-3 and can achieve excellent energy resolution of 7\% FWHM at 1 MeV\@.
However, the required large number of channels makes it an expensive choice and could lead to high radioactivity in the detector.
The alternative bar design has fewer readout channels, meaning much lower price and less radioactivity. 
Another advantage is its efficient $\gamma$ tagging capability. 
The drawbacks of the bar design are its worse resolution of 10\% FWHM at 1 MeV and possible ageing problems. 
A technology choice for the SuperNEMO calorimeter will be made by the end of 2009.

\section{Tracker R\&D}
The development programme of the tracker aims to optimise its operating parameters: wire length, material and configuration, detector readout and endcap design. 
Several small tracker prototypes were constructed and tested. 
A large 90-cell prototype (Fig.~\ref{fignemo}(b)) has undergone full commissioning of the detector and its readout, and is taking data with cosmics. 
Tracking studies have demonstrated efficient charge collection and plasma propagation along 4-metre long anode wires.
The achieved resolution is 0.7 mm in the transverse direction and 1.3 cm longitudinally.
In addition to the tracker R\&D, a dedicated wiring robot is being developed for mass production of 
drift cells.

\section{Source purity and the BiPo detector}
In order to achieve its target $0\nu\beta\beta$ half-life sensitivity, 
SuperNEMO must use isotope sources with high radiopurity.
A dedicated BiPo detector is being developed, 
which will be able to measure the purity of 10~m$^2$ of isotope source in a month with a sensitivity to ultra-low radioactive contaminations of $<2$~$\mu$Bq/kg for $^{208}$Tl and $<10$~$\mu$Bq/kg for $^{214}$Bi. 
The detector reconstructs the "BiPo" cascade of a $\beta$ decay followed by a delayed $\alpha$ decay, characteristic for  $^{208}$Tl and $^{214}$Bi.
The BiPo-1 prototype (Fig.~\ref{fignemo}(c)) consists of two thin sandwiched low-radioactive polystyrene plastic scintillators.
It has been running in LSM Modane since February 2008, measuring a background activity level of $A(^{208}$Tl$) = 1$~$\mu$Bq/m$^2$. 
A larger BiPo detector will be constructed in 2010. 


\begin{thebibliography}{99}

\bibitem{nemo}
R. Arnold et al., \emph{Technical design and performance of the NEMO 3 detector}, Nucl. Instr. Meth. {\bf A 536}, (2005) 79 [{\tt physics/0402115}]; 
R. Arnold et al., \emph{First results of the search of neutrinoless double beta decay with the NEMO 3 detector}, Phys. Rev. Lett. {\bf95}, (2005) 182302 [{\tt hep-ex/0507083}].

\bibitem{nme}
E. Caurier et al., \emph{The Influence of pairing on the nuclear matrix elements of the neutrinoless beta beta decays}, Phys. Rev. Lett. {\bf100}, (2008) 052503 [{\tt arXiv:0709.2137}]; 
F. Simkovic et al., \emph{Anatomy of nuclear matrix elements for neutrinoless double-beta decay}, Phys. Rev. {\bf C77} (2008) 045503 [{\tt arXiv:0710.2055}]; 
J. Suhonen et al., \emph{Nuclear matrix elements for double beta decay}, Int. J. Mod. Phys. {\bf E17} (2008) 1-11.

\end{thebibliography}
\end{document}